\documentclass[%
reprint,
superscriptaddress,
amsmath,amssymb,
aps,
]{revtex4-2}

\usepackage{graphicx}% Include figure files
\usepackage{dcolumn}% Align table columns on decimal point
\usepackage{bm}% bold math
\usepackage{multirow}
\usepackage{float}
\usepackage{array}
\usepackage{subfigure}
\usepackage{soul}
\usepackage{color}
\soulregister{\onlinecite}7
\soulregister{\cite}7
\soulregister{\ref}7
\sethlcolor{yellow}
\begin{document}

%\preprint{APS/123-QED}

\title{Constraining  Dark Matter Models with a Light Mediator from CDEX-10 Experiment at China Jinping Underground Laboratory}% Force line breaks with \\
% \thanks{A footnote to the article title}%

\author{Qi-Yuan Nie}%聂奇缘})}
\author{Wen-Han Dai}%代文翰})}
\author{Hao Ma}%马豪})}
 \email{mahao@tsinghua.edu.cn}

\author{Qian Yue}%岳骞})}
 \email{yueq@mail.tsinghua.edu.cn}
\author{Ke-Jun Kang}%康克军})}
\author{Yuan-Jing Li}%李元景})}
\affiliation{Key Laboratory of Particle and Radiation Imaging 
(Ministry of Education) and Department of Engineering Physics, 
Tsinghua University, Beijing 100084}

\author{Hai-Peng An}%安海鹏})}
\affiliation{Department of Physics, Tsinghua University, Beijing 100084}

\author{Greeshma C.}
\altaffiliation[Participating as a member of ]{TEXONO Collaboration}
\affiliation{Institute of Physics, Academia Sinica, Taipei 11529}
% \affiliation{Department of Physics, Central University of South Bihar, Gaya 824236}

\author{Jian-Ping Chang}%常建平})}
\affiliation{NUCTECH Company, Beijing 100084}

\author{Yun-Hua Chen}%陈云华})}
\affiliation{YaLong River Hydropower Development Company, Chengdu 610051}

\author{Jian-Ping Cheng}%程建平})}
\affiliation{Key Laboratory of Particle and Radiation Imaging 
(Ministry of Education) and Department of Engineering Physics, 
Tsinghua University, Beijing 100084}
\affiliation{School of Physics and Astronomy, Beijing Normal University, Beijing 100875}

\author{Zhi Deng}%邓智})}
\affiliation{Key Laboratory of Particle and Radiation Imaging 
(Ministry of Education) and Department of Engineering Physics, 
Tsinghua University, Beijing 100084}

\author{Chang-Hao Fang}%房昌昊})}
\affiliation{College of Physics, Sichuan University, Chengdu 610065}

\author{Xin-Ping Geng}%耿新平})}
\affiliation{Key Laboratory of Particle and Radiation Imaging 
(Ministry of Education) and Department of Engineering Physics, 
Tsinghua University, Beijing 100084}

\author{Hui Gong}%宫辉})}
\affiliation{Key Laboratory of Particle and Radiation Imaging 
(Ministry of Education) and Department of Engineering Physics, 
Tsinghua University, Beijing 100084}

\author{Tao Guo}%郭涛})}
\affiliation{Key Laboratory of Particle and Radiation Imaging 
(Ministry of Education) and Department of Engineering Physics, 
Tsinghua University, Beijing 100084}

\author{Xu-Yuan Guo}%郭绪元})}
\affiliation{YaLong River Hydropower Development Company, Chengdu 610051}

\author{Li He}%何力})}
\affiliation{Key Laboratory of Particle and Radiation Imaging 
(Ministry of Education) and Department of Engineering Physics, 
Tsinghua University, Beijing 100084}

\author{Jin-Rong He}%何金荣})}
\affiliation{YaLong River Hydropower Development Company, Chengdu 610051}

\author{Han-Xiong Huang}%黄瀚雄})}
\affiliation{Department of Nuclear Physics, China Institute of Atomic Energy, Beijing 102413}

\author{Tu-Chen Huang}%黄土琛})}
\affiliation{Sino-French Institute of Nuclear and Technology, Sun Yat-sen University, Zhuhai 519082}

\author{S. Karmakar}
\altaffiliation[Participating as a member of ]{TEXONO Collaboration}
\affiliation{Institute of Physics, Academia Sinica, Taipei 11529}
% \affiliation{Department of Physics, GLA University, Mathura 281406}

\author{Jian-Min Li}%李荐民})}
\author{Jin Li}%李金})}
\affiliation{Key Laboratory of Particle and Radiation Imaging 
(Ministry of Education) and Department of Engineering Physics, 
Tsinghua University, Beijing 100084}

\author{Yu-Lan Li}%李玉兰})}
\affiliation{Key Laboratory of Particle and Radiation Imaging 
(Ministry of Education) and Department of Engineering Physics, 
Tsinghua University, Beijing 100084}

\author{Hau-Bin Li}%李浩斌})}
\altaffiliation[Participating as a member of ]{TEXONO Collaboration}
\affiliation{Institute of Physics, Academia Sinica, Taipei 11529}

\author{Ming-Chuan Li}%李名川})}
\affiliation{YaLong River Hydropower Development Company, Chengdu 610051}

\author{Han-Yu Li}%李含宇})}
\affiliation{College of Physics, Sichuan University, Chengdu 610065}

\author{Qian-Yun Li}%李倩沄})}
\author{Ren-Ming-Jie Li}%李任明杰})}
\affiliation{College of Physics, Sichuan University, Chengdu 610065}

\author{Xue-Qian Li}%李学潜})}
\affiliation{School of Physics, Nankai University, Tianjin 300071}

\author{Yi-Fan Liang}%梁艺帆})}
\affiliation{Key Laboratory of Particle and Radiation Imaging 
(Ministry of Education) and Department of Engineering Physics, 
Tsinghua University, Beijing 100084}

\author{Bin Liao}%廖斌})}
\affiliation{School of Physics and Astronomy, Beijing Normal University, Beijing 100875}

\author{Fong-Kay Lin}%林枫凯})}
\altaffiliation[Participating as a member of ]{TEXONO Collaboration}
\affiliation{Institute of Physics, Academia Sinica, Taipei 11529}

\author{Shin-Ted Lin}%林兴德})}
\affiliation{College of Physics, Sichuan University, Chengdu 610065}

\author{Jia-Xuan Liu}%刘家璇})}
\affiliation{Key Laboratory of Particle and Radiation Imaging 
(Ministry of Education) and Department of Engineering Physics, 
Tsinghua University, Beijing 100084}

\author{Yan-Dong Liu}%刘言东})}
\affiliation{School of Physics and Astronomy, Beijing Normal University, Beijing 100875}

\author{Yuan-Yuan Liu}%刘圆圆})}
\affiliation{School of Physics and Astronomy, Beijing Normal University, Beijing 100875}

\author{Shu-Kui Liu}%刘书魁})}
\affiliation{College of Physics, Sichuan University, Chengdu 610065}

\author{Yu Liu}%刘钰})}
\affiliation{College of Physics, Sichuan University, Chengdu 610065}

\author{Hui Pan}%潘辉})}
\affiliation{NUCTECH Company, Beijing 100084}

\author{Ning-Chun Qi}%祁宁春})}
\affiliation{YaLong River Hydropower Development Company, Chengdu 610051}

\author{Jie Ren}%任杰})}
\author{Xi-Chao Ruan}%阮锡超})}
\affiliation{Department of Nuclear Physics, China Institute of Atomic Energy, Beijing 102413}

\author{Man-Bin Shen}%申满斌})}
\affiliation{YaLong River Hydropower Development Company, Chengdu 610051}

\author{Manoj Kumar Singh}
\altaffiliation[Participating as a member of ]{TEXONO Collaboration}
\affiliation{Institute of Physics, Academia Sinica, Taipei 11529}
\affiliation{Department of Physics, Banaras Hindu University, Varanasi 221005}

\author{Wen-Liang Sun}%孙文良})}
\affiliation{YaLong River Hydropower Development Company, Chengdu 610051}

\author{Tian-Xi Sun}%孙天希})}
\affiliation{School of Physics and Astronomy, Beijing Normal University, Beijing 100875}

\author{Chang-Jian Tang}%唐昌建})}
\affiliation{College of Physics, Sichuan University, Chengdu 610065}

\author{Yang Tian}%田阳})}
\affiliation{Key Laboratory of Particle and Radiation Imaging 
(Ministry of Education) and Department of Engineering Physics, 
Tsinghua University, Beijing 100084}

\author{Hong-Fei Wan}%万宏飞})}
\author{Jun-Zheng Wang}%王军正})}
\affiliation{Key Laboratory of Particle and Radiation Imaging 
(Ministry of Education) and Department of Engineering Physics, 
Tsinghua University, Beijing 100084}

\author{Yu-Feng Wang}%王钰锋})}
\affiliation{Key Laboratory of Particle and Radiation Imaging 
(Ministry of Education) and Department of Engineering Physics, 
Tsinghua University, Beijing 100084}

\author{Guang-Fu Wang}%王广甫})}
\affiliation{School of Physics and Astronomy, Beijing Normal University, Beijing 100875}

\author{Li Wang}%王力})}
\affiliation{School of Physics and Astronomy, Beijing Normal University, Beijing 100875}

\author{Qing Wang}%王青})}
\affiliation{Key Laboratory of Particle and Radiation Imaging 
(Ministry of Education) and Department of Engineering Physics, 
Tsinghua University, Beijing 100084}
\affiliation{Department of Physics, Tsinghua University, Beijing 100084}

\author{Henry-Tsz-King Wong}%王子敬})}
\altaffiliation[Participating as a member of ]{TEXONO Collaboration}
\affiliation{Institute of Physics, Academia Sinica, Taipei 11529}

\author{Yu-Cheng Wu}%吴玉成})}
\affiliation{Key Laboratory of Particle and Radiation Imaging 
(Ministry of Education) and Department of Engineering Physics, 
Tsinghua University, Beijing 100084}

\author{Hao-Yang Xing}%幸浩洋})}
\affiliation{College of Physics, Sichuan University, Chengdu 610065}

\author{Kai-Zhi Xiong}%熊开智})}
\affiliation{YaLong River Hydropower Development Company, Chengdu 610051}

\author{Rui Xu}%徐锐})}
\affiliation{Key Laboratory of Particle and Radiation Imaging 
(Ministry of Education) and Department of Engineering Physics, 
Tsinghua University, Beijing 100084}

\author{Yin Xu}%徐音})}
\affiliation{School of Physics, Nankai University, Tianjin 300071}

\author{Tao Xue}%薛涛})}
\affiliation{Key Laboratory of Particle and Radiation Imaging 
(Ministry of Education) and Department of Engineering Physics, 
Tsinghua University, Beijing 100084}

\author{Yu-Lu Yan}%鄢雨璐})}
\affiliation{College of Physics, Sichuan University, Chengdu 610065}

\author{Li-Tao Yang}%杨丽桃})}
\affiliation{Key Laboratory of Particle and Radiation Imaging 
(Ministry of Education) and Department of Engineering Physics, 
Tsinghua University, Beijing 100084}

\author{Nan Yi}%易难})}
\affiliation{Key Laboratory of Particle and Radiation Imaging 
(Ministry of Education) and Department of Engineering Physics, 
Tsinghua University, Beijing 100084}

\author{Chun-Xu Yu}%喻纯旭})}
\affiliation{School of Physics, Nankai University, Tianjin 300071}

\author{Hai-Jun Yu}%于海军})}
\affiliation{NUCTECH Company, Beijing 100084}

\author{Xiao Yu}%于骁})}
\author{Ming Zeng}%曾鸣})}
\author{Zhi Zeng}%曾志})}
\author{Zhen-Hua Zhang}%张振华})}
\author{Zhen-Yu Zhang}%张震宇})}
\author{Peng Zhang}%张鹏})}
\affiliation{Key Laboratory of Particle and Radiation Imaging 
(Ministry of Education) and Department of Engineering Physics, 
Tsinghua University, Beijing 100084}

\author{Feng-Shou Zhang}%张丰收})}
\affiliation{School of Physics and Astronomy, Beijing Normal University, Beijing 100875}

\author{Lei Zhang}%张磊})}
\affiliation{College of Physics, Sichuan University, Chengdu 610065}

\author{Ji-Zhong Zhao}%赵纪仲})}
\affiliation{Key Laboratory of Particle and Radiation Imaging 
(Ministry of Education) and Department of Engineering Physics, 
Tsinghua University, Beijing 100084}

\author{Kang-Kang Zhao}%赵康康})}
\affiliation{College of Physics, Sichuan University, Chengdu 610065}

\author{Ming-Gang Zhao}%赵明刚})}
\affiliation{School of Physics, Nankai University, Tianjin 300071}

\author{Ji-Fang Zhou}%周济芳})}
\affiliation{YaLong River Hydropower Development Company, Chengdu 610051}

\author{Zu-Ying Zhou}%周祖英})}
\affiliation{Department of Nuclear Physics, China Institute of Atomic Energy, Beijing 102413}

\author{Jing-Jun Zhu}%朱敬军})}
\affiliation{College of Physics, Sichuan University, Chengdu 610065}

\collaboration{CDEX Collaboration}%\noaffiliation

%\date{\today}% It is always \today, today,
            %  but any date may be explicitly specified

\begin{abstract}
We search for nuclear recoil signals of dark matter models with a light mediator using data taken from a p-type point-contact germanium detector of the CDEX-10 experiment at the China Jinping Underground Laboratory.
The 90\% confidence level upper limits on the DM-nucleon interaction cross section from 205.4 kg-day exposure data are derived, excluding new parameter space in 2$\sim$5 GeV DM mass when the mediator mass is comparable to or lighter than the typical momentum transfer.
We further interpret our results to constrain a specific self-interacting dark matter model with a light mediator coupling to the photon through kinetic mixing, and set experimental limits on the model parameter region favored by astrophysical observations.
\\
\\
\textbf{Key words:} dark matter, direct detection, light mediator, self-interacting dark matter
\end{abstract}

\keywords{dark matter, direct detection, self-interacting dark matter}

\maketitle

\section{Introduction}\label{sec.I}
Various cosmological and astronomical observations provide compelling evidence of the existence of dark matter (DM) in the Universe~\cite{b1}. 
The weakly interacting massive particle (WIMP) is a well-motivated DM candidate that could naturally explain the relic abundance of DM~\cite{b2,b3,b4}, while decades of direct detection experiments have not found positive signal yet~\cite{b5,b6,b7,b8,b9}. 
However, introducing a light force carrier which mediates the DM-nucleus interaction can expand the regions of direct detection experiments. 

If a mediator couples to both DM particles and standard model (SM) particles, the DM particle could interact with nucleus via the exchange of the mediator. 
In this case, the nuclear recoil spectrum depends sensitively on both the DM mass and mediator mass. 
Furthermore, the smallness of the mediator mass leads to a huge enhancement of direct detection cross sections, so that an observation of DM scattering may be possible in spite of the small couplings~\cite{b10}.
Light mediators have been widely proposed in many different DM models, such as hidden sector models~\cite{b11}, light DM models~\cite{b12} and self-interacting DM (SIDM) models~\cite{b13,b14}. 

SIDM models where DM particles scatter elastically with each other are partly motivated to solve the small-scale challenges to the $\Lambda$CDM Paradigm~\cite{b14,b15,b16}. DM self-interaction can change the inner structure of DM halos and better explain astrophysical observations in galaxies, while keeping all the success of $\Lambda$CDM on larger scales.
To have an observable effect on DM halos over cosmological time scales, the cross section of DM self-scattering per unit mass must be of order~\cite{b13}
\begin{equation}
\label{eq1}
\sigma/m_\chi \sim \mathrm{1 \; cm^2/g} \approx \mathrm{2 \times 10^{-24} \; cm^2/GeV}
\end{equation}
where $m_\chi$ is the DM particle mass.
This value of $\sigma/m_\chi$, as shown in Eq.~\eqref{eq1}, significantly exceeds the expectation from weak-scale physics. 
For a typical WIMP, the cross section per unit mass is $\sigma/m_\chi$$\sim$10$^{-38}$\;$\mathrm {cm^2/GeV}$.
Naturally, a light mediator is introduced to participate in the DM self-interaction~\cite{b17}.
The mediator mass is comparable to or lighter than the typical momentum transfer ($\mathcal{O}$(10) MeV) to yield the required cross section of DM self-scattering.
Coupling with SM particles, light mediator particles have been directly searched for SIDM models~\cite{b18,b19,b20}. 

In this work, we report the limits on the zero-momentum DM-nucleon interaction cross section through a light mediator based on the 205.4 kg-day exposure data from the CDEX-10 experiment. 
We further apply our results to constrain a well-motivated SIDM model with a light mediator.   

\section{Experiment and Data Analysis}\label{sec.II}
The CDEX-10 experiment, aiming at the search of light DM, operates a 10-kg p-type point-contact germanium (PPCGe) detector array in the China Jinping Underground Laboratory (CJPL) with a rock overburden of about 2400 meters~\cite{b21}. The detector array, consisting of three triple-element PPCGe detector strings (C10-A, B, C) encapsulated in copper vacuum tubes, is directly immersed in liquid nitrogen (LN$_2$) for cooling. A passive shield composed of 20 cm thick high-purity oxygen-free copper surrounds the detector array in the LN$_2$ cryostat against environmental radioactivity. The LN$_2$ cryostat and data acquisition (DAQ) system operate in a shielding room with 1 m thick polyethylene walls at CJPL-I. The detailed configuration of the CDEX-10 experiment was described in our previous works~\cite{b22}.

A 205.4 kg-day dataset from C10-B1 in the nine detectors is used for DM search in this work. C10-B1, with the dead layer thicknesses of 0.88$\pm$0.12 mm and a fiducial mass of 939 g, was operated for data taking from February 2017 to August 2018. The data analysis followed several procedures described in previous papers~\cite{b22,b23}, including pedestal cut, physics-noise events cut, and bulk or surface events cut. The dead time of the DAQ system caused by the reset of the charge-sensitive preamplifier was measured to be 5.7\% by random trigger events. The detector achieved an analysis threshold of 160 eVee (“eVee” represents electron equivalent energy) with the combined efficiency being 4.5\% and the event rate being 2.5 counts $\mathrm{kg^{-1} \;keV^{-1} \;day^{-1}}$ in the 2$\sim$4 keV energy range after all event selections and efficiency correction~\cite{b22,b24}.

\begin{figure}[htbp]
\includegraphics[width=\hsize]{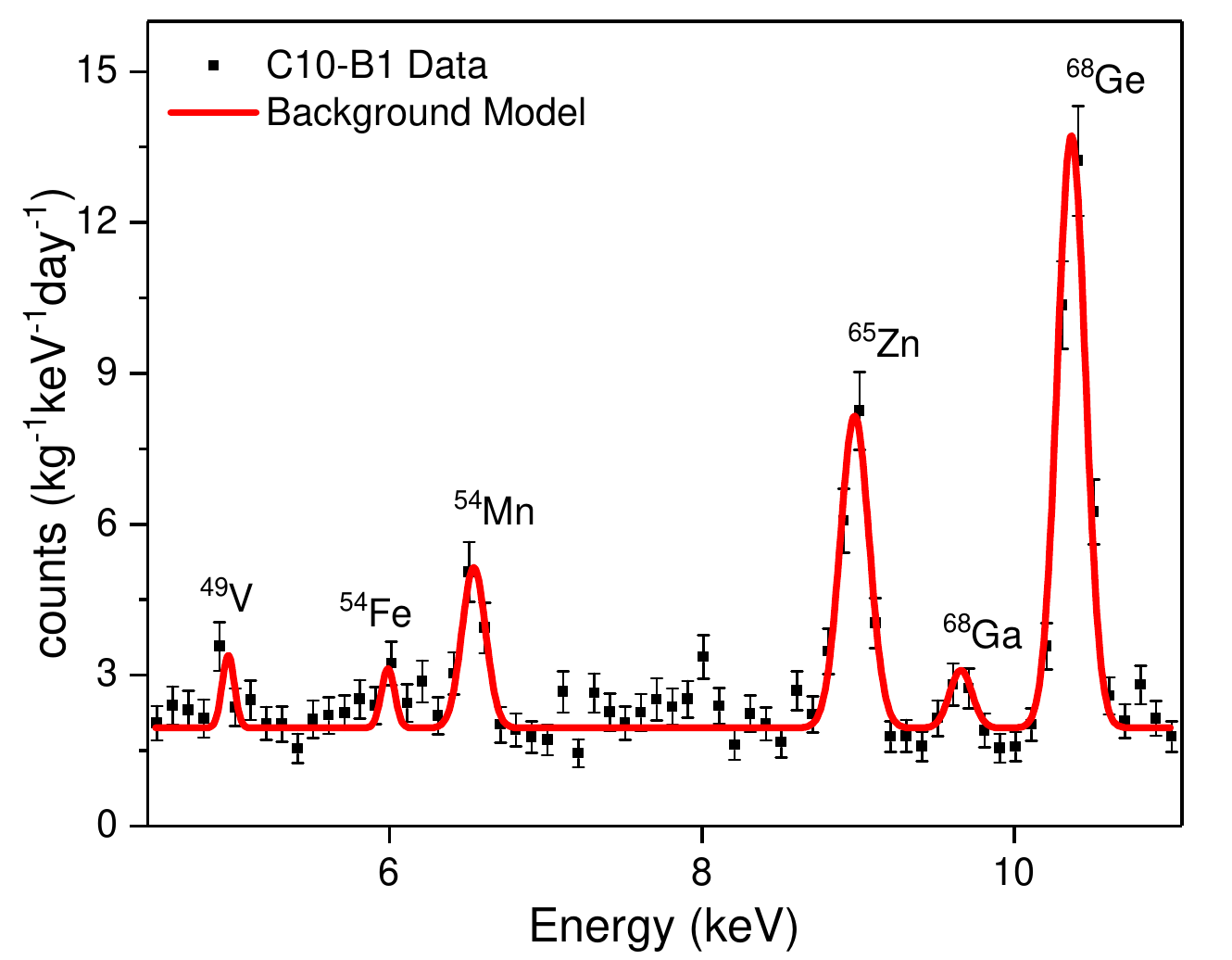}
\caption{Spectrum fitting of the K-shell X-ray peaks of cosmogenic radionuclides in C10-B1 data. The fitting energy range is from 4.5 to 11 keV, with an energy bin width of 100 eV.}
\label{fig1}
\end{figure}

A minimal-$\chi^2$ analysis method~\cite{b8} is applied to the C10-B1 energy spectrum from 0.16 to 2.16 keV to search for nuclear recoil signals of dark matter. The $\chi^2$ is defined as
\begin{equation}
\label{eq2}
\chi^2 = \sum_{i}{\frac{[n_i-S_i(m_\chi, m_\phi, \sigma_{\chi N})-B_i]^2}{\sigma_{\mathrm {stat},i}^2+\sigma_{\mathrm {syst},i}^2}}
\end{equation}
where $n_i$ is the measured event rate at the $i^{th}$ energy bin, and $S_i(m_\chi, m_\phi, \sigma_{\chi N})$ is the expected DM event rate corresponding to the DM mass $m_\chi$, mediator mass $m_\phi$ and DM-nucleus interaction cross section $\sigma_{\chi N}$. Uncertainty $\sigma_{\mathrm {stat},i}$ and $\sigma_{\mathrm {syst},i}$ represent the statistical and systematical components. $B_i$ denotes the background event rate. The background contribution of C10-B1 is assumed to be a flat continuum ($p_{\mathrm 0}$) plus known M-shell and L-shell X-ray peaks~\cite{b8} and can be constructed as
\begin{equation}
\label{eq3}
B = p_{\mathrm 0} + \sum{I \cdot \frac{1}{\sqrt{2\pi}\sigma}\,\mathrm{exp}\,(-{\frac{(E-E_{\mathrm {M/L}})^2}{2\sigma^2}})}
\end{equation}
where $E_{\mathrm {M/L}}$ is the peak energy of the M/L-shell X-rays listed in Tab.~\ref{tab1}, and $\sigma$ is the energy resolution. $I$ is the intensity of the M/L-shell peak and is constrained by the intensity of the corresponding K-shell peak, which was measured as shown in Fig.~\ref{fig1}.

The best estimator of the DM-nucleon interaction cross section at a certain mediator mass is probed by $\chi^2$ minimization of the right hand-side of Eq.~\eqref{eq2} with the background model shown in Eq.~\eqref{eq3}. 
Upper limits at the 90\% confidence level (C.L.) are computed following the Feldman-Cousins method~\cite{b25}.
\begin{table}[htbp]
\centering
\caption{Characteristic X-rays from cosmogenic radionuclides in germanium~\cite{b8,b26}.}
\label{tab1}
    \centering
    \resizebox{0.45\textwidth}{!}{
    \begin{tabular}{lcccc}
    \hline
    \multirow{2}{*}{Nuclide} &
    \multirow{2}{*}{$T_{1/2}$} &
    \multicolumn{2}{c}{X-ray energy (keV)} &
    \multirow{2}{*}{K/L ratio} \\
    \cline{3-4}
    & & K-shell & M/L-shell & \\
    \hline
    $^{68}$Ge &270.9 d & 10.37 & 1.30(L) & 0.12 \\
              &        &       & 0.16(M) & 0.03 \\
    $^{68}$Ga & 68.1 min & 9.66 & 1.19(L) & 0.11 \\ 
    $^{65}$Zn &243.9 d & 8.98 & 1.10(L) & 0.12 \\ 
    $^{55}$Fe &2.7 yr & 6.54 & 0.76(L) & 0.11 \\ 
    $^{54}$Mn &312.2 d & 5.99 & 0.70(L) & 0.11 \\ 
    $^{49}$V &330.0 d & 4.97 & 0.56(L) & 0.11 \\ 
    \hline
    \end{tabular}
    }
\end{table}

\section{Constraining dark matter models with a light mediator}\label{sec.III}
We first explore a general scenario where the interaction between DM and nucleon is mediated by a force carrier, $\phi$. The light mediator is further assumed to possess equal effective couplings to the proton and neutron, motivated by the standard WIMP search. The cross section of DM–nucleon scattering via a light mediator can be expressed as~\cite{b13}
\begin{equation}
\label{eq4}
\sigma_{\chi N} = \sigma(q^2=0)A^2(\frac{\mu}{\mu_p})^2\;\frac{m_{\phi}^4}{(m_{\phi}^2+q^2)^2}F^2(q^2)
\end{equation}
where $\sigma(q^2=0)$ is the DM-nucleon interaction cross section with zero momentum transfer $(q^2=0)$, $A$ is the mass number of target nucleus, $\mu(\mu_p)$ is the DM-nucleus (nucleon) reduced mass, $m_\phi$ is the mediator mass, and $F(q^2)$ is the Helm form factor of target nucleus~\cite{b27}. $\sigma_{\chi N}$ is momentum dependent and negatively correlated with momentum transfer, which benefits detectors with lighter target nucleus. When $m_\phi$\,$\gg$\,$q$, $\sigma_{\chi N}$ converges to the standard WIMP case. 

The expected energy spectrum of DM signal is described as~\cite{b28}
\begin{equation}
\label{eq5}
\frac{dR}{dE} =\frac{\sigma_{\chi N}\rho_{\chi}}{2m_{\chi}{\mu}^2}\;\int_{v_{\mathrm {min}}(E_R)}\frac{f(v,t)}{v}d^3v
\end{equation}
where the local DM density $\rho_{\chi}$ is set to 0.3 $\mathrm {GeV/cm^3}$, $v_{\mathrm {min}}(E_R)$ is the minimum DM velocity at a given recoil energy $E_R$, and $f(v,t)$ is the time-dependent DM velocity distribution relative to the detector~\cite{b29}.

\begin{figure}[htbp]
\includegraphics[width=\hsize]{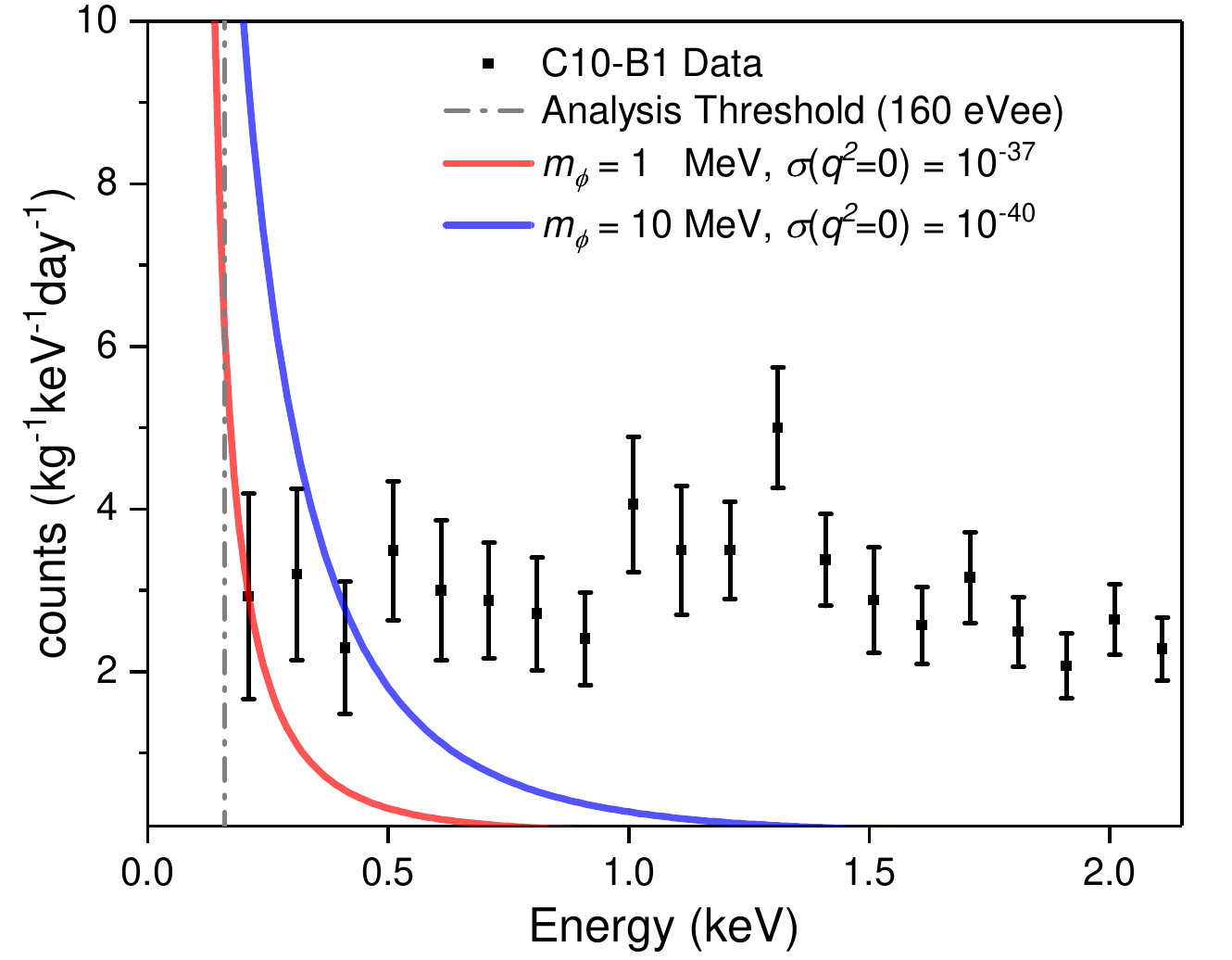}
\caption{Expected spectra in the C10-B1 detector for 10 GeV DM particles with $m_\phi=1$ MeV (red) and $m_\phi=10$ MeV (blue).
The energy resolution is fitted by the zero energy (defined by the random trigger events) and the cosmogenic X-ray peaks as shown in Fig.~\ref{fig1}. The black dot represents the C10-B1 spectrum from 0.16 to 2.16 keV, with an energy bin width of 100 eV. The gray dash indicates the analysis threshold.}
\label{fig2}
\end{figure}

The Fig.~\ref{fig2} shows the expected nuclear recoil spectra in the C10-B1 detector for 10 GeV DM particles with $m_\phi=1$ MeV (red) and $m_\phi=10$ MeV (blue), calculated by Eq.~\eqref{eq4} and Eq.~\eqref{eq5}.
These spectra fall steeply in the low-energy region and also affected by the mediator mass.
Thus, low background detectors with low energy thresholds, like PPCGe detectors, are advantageous to search for this type of DM signals.

\begin{figure}[htbp]
\includegraphics[width=\hsize]{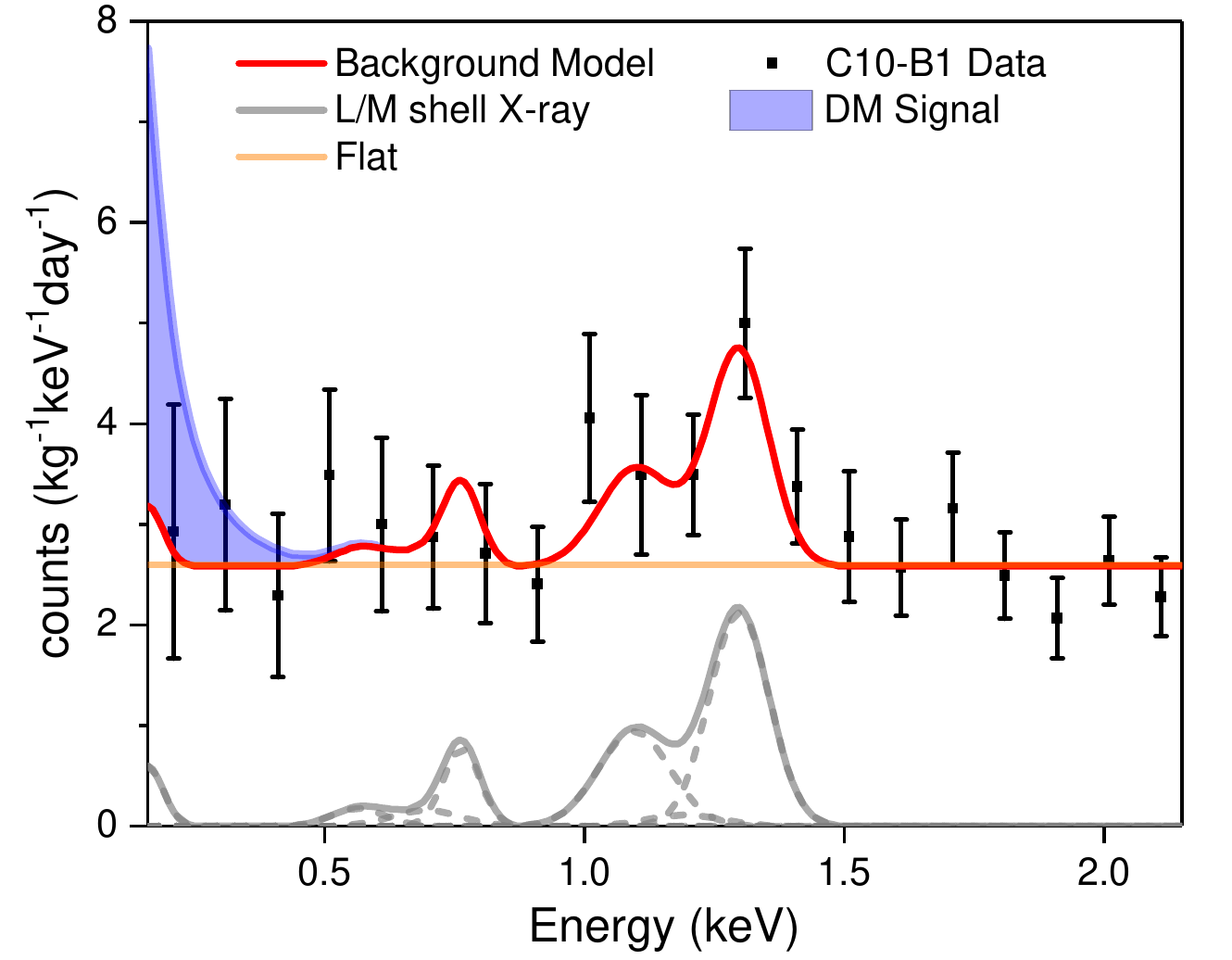}
\caption{The best-fit spectrum with the DM signal for ($m_\chi$, $m_\phi$) = (5, 0.01) GeV. The background model (red) consists of the flat part (orange) and the contributions from X-ray peaks (gray). The DM signal is depicted in blue.}
\label{fig3}
\end{figure}

\begin{figure}[htbp]
\includegraphics[width=\hsize]{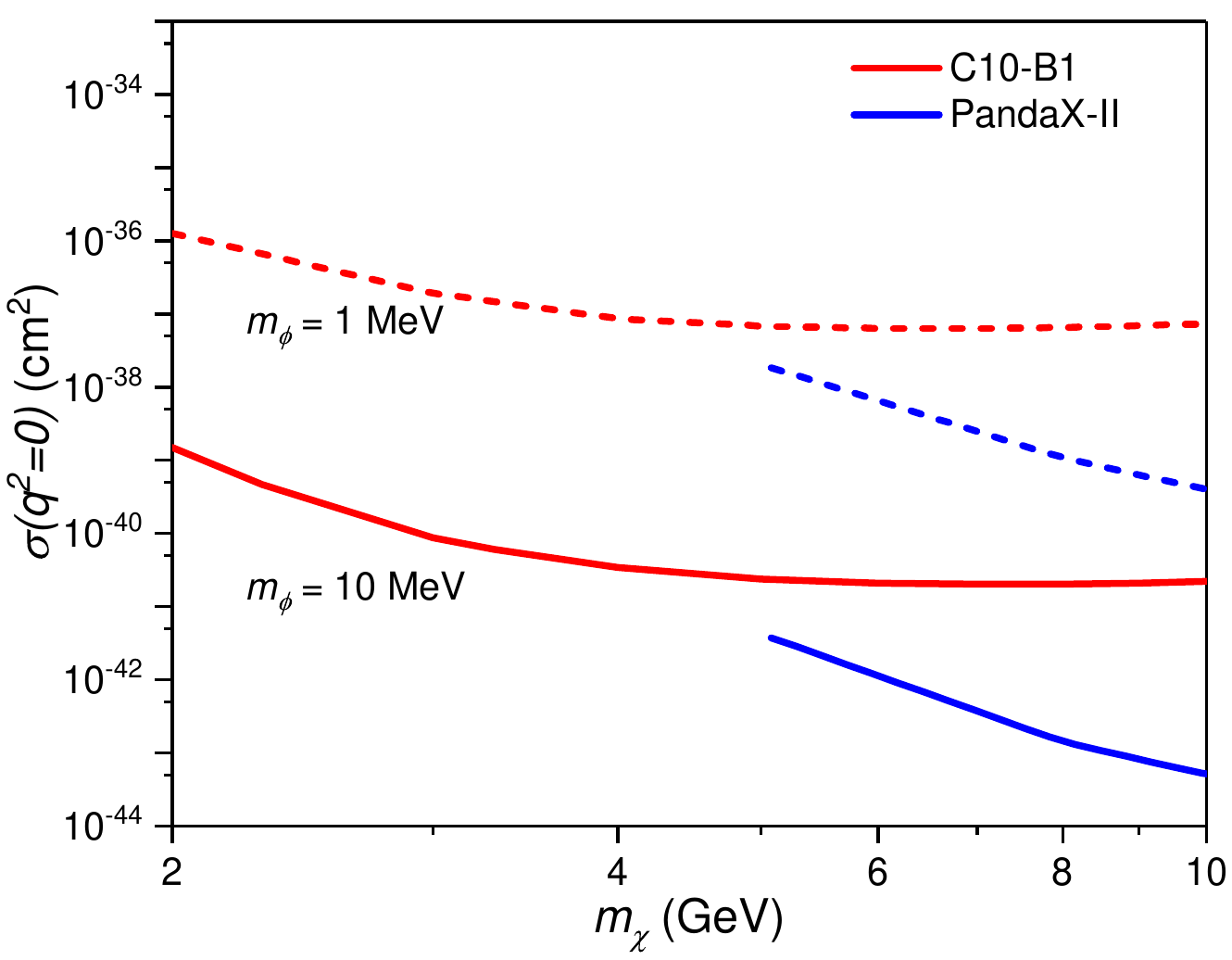}
\caption{C10-B1 90\% C.L. upper limits (red) on the DM-nucleon interaction cross section for DM models with mediator mass $m_\phi$ = 1 MeV (dash line) and 10 MeV (solid line). The limits from PandaX-II~\cite{b20} are also shown for comparison (blue).}
\label{fig4}
\end{figure}

We scanned the DM mass in the range of 2$\sim$10 GeV for $m_\phi$ = 1 MeV and 10 MeV, and no significant signal was observed. 
The best-fit spectrum with the DM signal for ($m_\chi$, $m_\phi$) = (5, 0.01) GeV is shown in Fig.~\ref{fig3}.
By minimizing the $\chi^2$ values, the upper limits at 90\% C.L. were derived and shown in Fig.~\ref{fig4}.
For $m_\phi$ = 1 MeV, the upper limit is $\sigma(q^2=0) = \mathrm{1.6\times10^{-36} cm^2}$ at $m_\chi$ = 2 GeV. 
For $m_\phi$ = 10 MeV, the upper limit is $\sigma(q^2=0) = \mathrm{3.7\times10^{-39} cm^2}$ at $m_\chi$ = 2 GeV. 
Compared with the results from PandaX-II~\cite{b20}, this work used a high-purity germanium detector with lower energy threshold and excluded new parameter space in lower DM mass ranging from 2 to 5 GeV.

We consider a specific case, where DM is assumed to be a Dirac fermion and it couples to a light vector mediator with a gauge mixing~\cite{b13,b20}. 
We further assume the mediator couples to photon through kinetic mixing~\cite{b31}.
The DM-nucleon interaction cross section with zero momentum transfer $(q^2=0)$ can be written as~\cite{b20}:
\begin{equation}
\label{eq6}
\sigma(q^2=0) = \frac{16\pi\alpha_{\mathrm {EM}}\alpha_{\chi}\mu_p^2}{m_{\phi}^4}\;\left [\frac{\epsilon_{\gamma}Z}{A}\right ]^2
\end{equation}
where $\alpha_{\mathrm {EM}}$ and $\alpha_{\chi}$ are the fine structure constants in the visible and dark sectors. We set $\alpha_{EM}$ = 1/137 and take the value of $\alpha_{\chi}$ from  Ref.~\cite{b30}. $\epsilon_{\gamma}$ is the kinetic mixing parameter, and $Z$ is the proton number of target nucleus. 

\begin{figure}[htbp]
\subfigure{
\label{fig5a}
\includegraphics[width=\hsize]{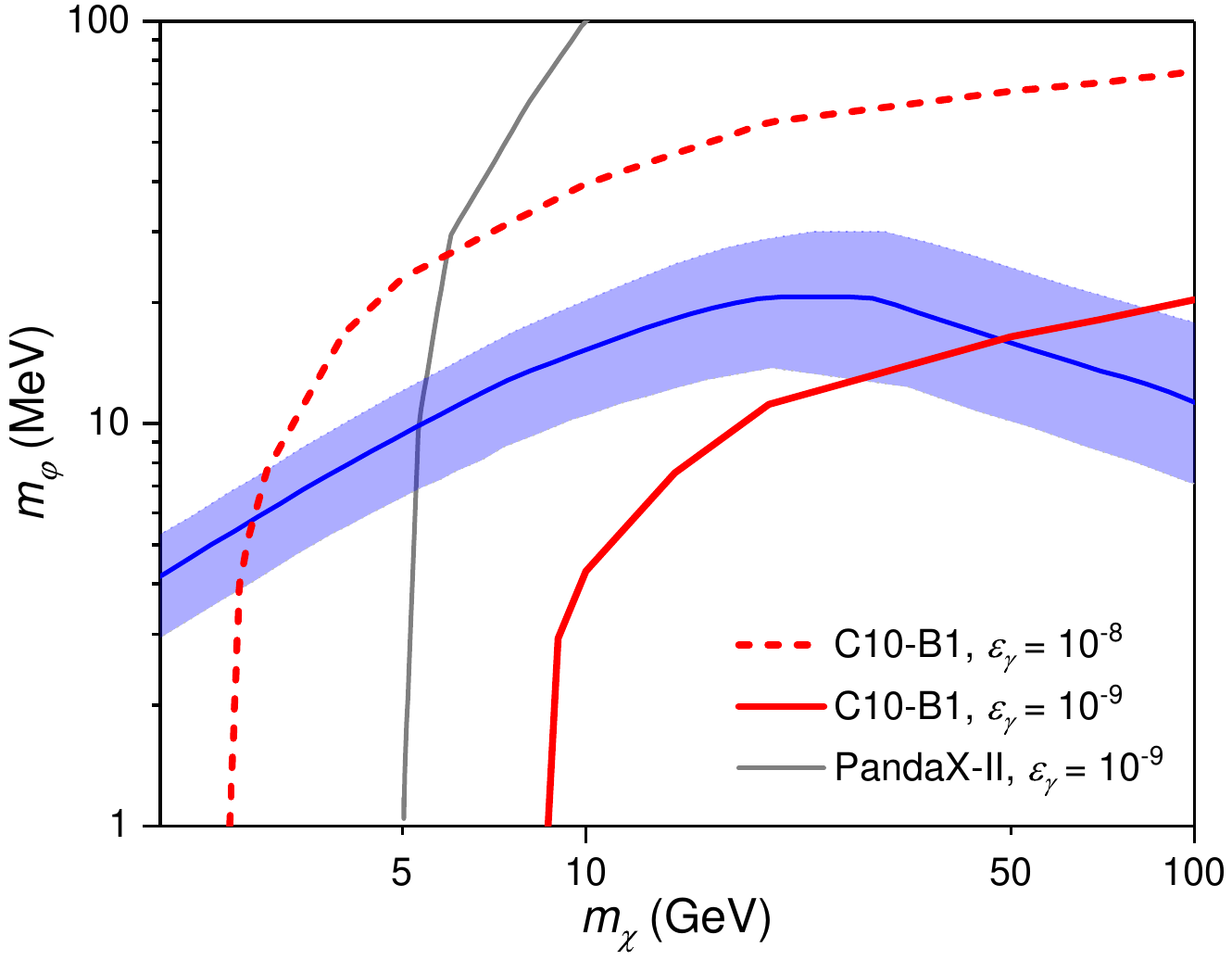}
}
\subfigure{
\label{fig5b}
\includegraphics[width=\hsize]{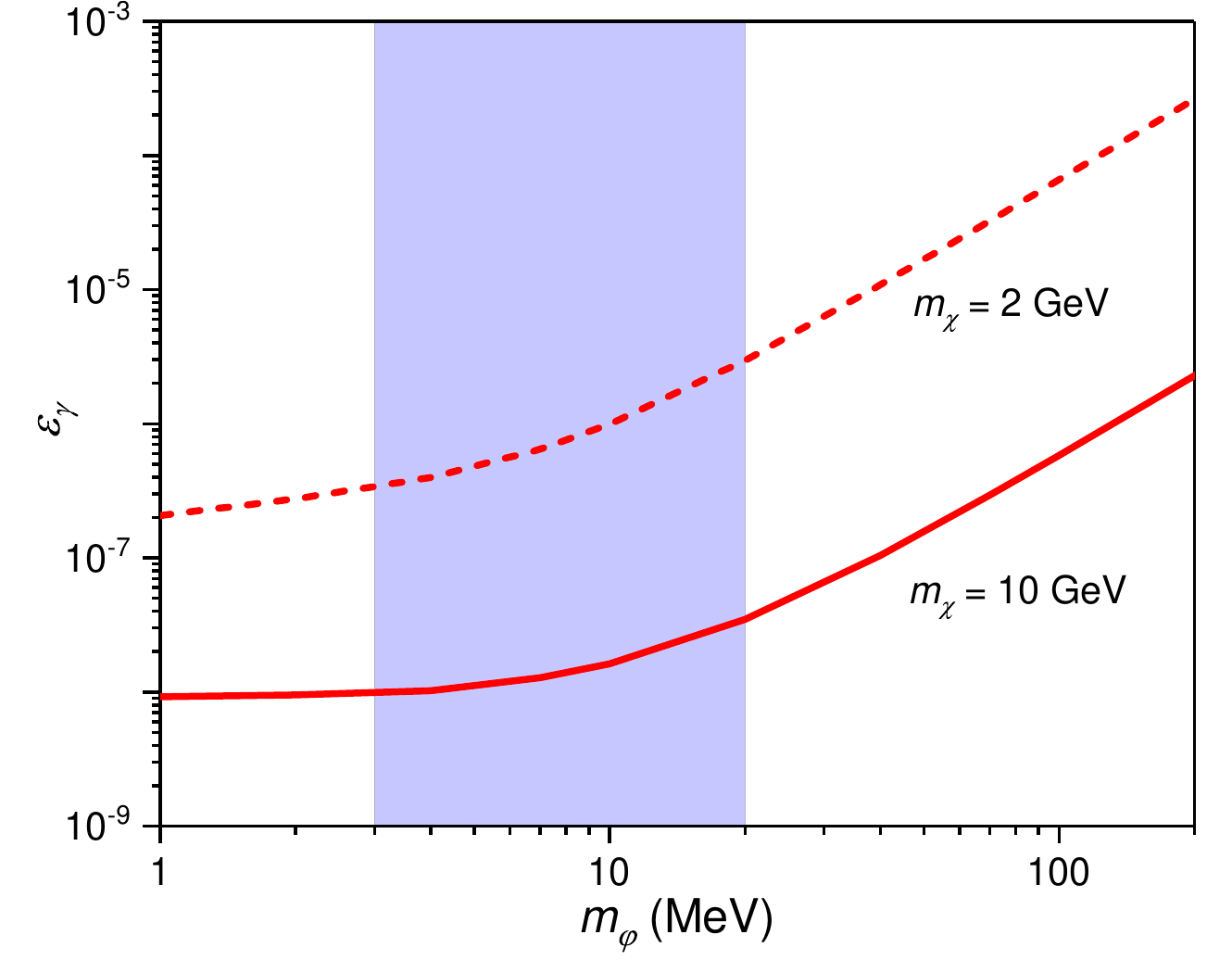}
}
\caption{\emph{Top:} CDEX-10 90\% C.L. lower limits (red) on the mediator mass for the kinetic mixing parameter $\epsilon_{\gamma}$ = 10$^{-8}$ (dash line) and 10$^{-9}$ (solid line). 
The blue band indicates the SIDM parameter region favored by astrophysical observations of DM halos~\cite{b30}. 
The limits from PandaX-II~\cite{b20} are also shown for comparison (gray solid line).
\emph{Bottom:} CDEX-10 90\% C.L. upper limits (red) on the kinetic mixing parameter for the mediator mass $m_\chi$ = 2 GeV (dash line) and 10 GeV (solid line). The vertical band indicates the mediator mass range for SIDM.}
\label{fig5}
\end{figure}

The 90\% C.L. lower limits in the $m_\phi$-$m_\chi$ plane for two $\epsilon_{\gamma}$ values ($\mathrm{10^{-8}}$ and $\mathrm{10^{-9}}$) of Eq.~\eqref{eq6} are derived.
Our results can constrain a large portion of the SIDM parameter space favored by astrophysical observations. Shown in the top panel of Fig.~\ref{fig5}, the sensitivity improves as $\epsilon_{\gamma}$ increases. 
For $\epsilon_{\gamma} = \mathrm{10^{-8}}$, the region with $m_\chi$ > 2 GeV is excluded.
For $\epsilon_{\gamma} = \mathrm{10^{-9}}$, the region with $m_\chi$ > 50 GeV is excluded.
Alternatively, we can derive upper limits in the $\epsilon_{\gamma}$-$m_{\chi}$ plane for given $m_\chi$, shown in the bottom panel of Fig.~\ref{fig5} with $m_\chi$ = 2 and 10 GeV for instance.

\section{Summary} \label{sec.V}

We have used the 205.4 kg-day dataset from CDEX-10 experiment to constrain dark matter models with a light mediator.
With an analysis threshold of 160 eVee, we set new limits on DM-nucleon interaction cross section for the DM mass ranging from 2 to 5 GeV when the mediator mass is comparable to or less than the typical momentum transfer ($\mathcal{O}$(10) MeV). We further apply our results to constrain the parameter space under a SIDM model with a light mediator (photon) mixing with SM particles. 
With the kinetic mixing parameter of $\mathrm{10^{-8}}$, we excluded the parameter region favored by astrophysical observations with $m_\chi$ > 2 GeV.
Upper limits on the kinetic mixing parameter for DM mass of 2 GeV and 10 GeV are also derived.

\begin{acknowledgments}
    \setlength{\parskip}{0.0cm}
    This work was supported by the National Key Research 
    and Development Program of China 
    (Grant No. 2023YFA1607101) and 
    the National Natural Science Foundation of China 
    (Grants No. 12425507, 12322511, 12175112).
    We would like to thank CJPL and its staff for hosting and supporting the CDEX project. 
    CJPL is jointly operated by Tsinghua University and Yalong River Hydropower Development Company.
\end{acknowledgments}

% The \nocite command causes all entries in a bibliography to be printed out
% whether or not they are actually referenced in the text. This is appropriate
% for the sample file to show the different styles of references, but authors
% most likely will not want to use it.
\nocite{*}

% \bibliography{PaperRef}% Produces the bibliography via BibTeX.
% \bibliographystyle{plainnat}

\end{document}